\documentclass[prb,twocolumn,showpacs,amsmath,amssymb]{revtex4-1}
\usepackage{graphicx} 
\usepackage{epsfig}
\usepackage{dcolumn} 
\usepackage{color}
\newcommand{\be}{\begin{equation}}
\newcommand{\ee}{\end{equation}}
\newcommand{\bea}{\begin{eqnarray}}
\newcommand{\eea}{\end{eqnarray}}

\renewcommand{\vec}[1]{{\bf #1}}

\newcommand{\qvec}{{\bf q}}   

\newcommand{\kvec}{{\bf k}}

\begin{document}
\title{Effect of the Anomalous Diffusion of Fluctuating Cooper Pairs in the Density of States of Superconducting 
NbN Thin Films}
\author{Pietro Brighi$^1$, Marco Grilli$^{2,3,*}$, Brigitte Leridon$^4$, Sergio Caprara$^{2,3}$ }
\affiliation{$^1$ ST Austria, Am Campus 1, 3400, Klosterneuburg, Austria \\
$^2$ ISC-CNR, via dei Taurini 19, I-00185 Rome, Italy \\
$^3$ Department of Physics, Universit\`a  La Sapienza, Piazzale A. Moro 5, I-00185 Rome, Italy\\
$^4$ LPEM, ESPCI Paris, CNRS, Universit\'e PSL, Sorbonne Universit\'es, 10 rue Vauquelin, 75005 Paris, France\\
$^*$ Correspondence should be addressed to M.G.(marco.grilli@roma1.infn.it)}

\begin{abstract}
Recent scanning tunnelling microscopy experiments in NbN thin disordered superconducting films found an 
emergent inhomogeneity at the scale of tens of nanometers. This inhomogeneity is mirrored by an apparent 
dimensional crossover in the paraconductivity measured in transport above the superconducting critical temperature 
$T_c$. This behavior was interpreted in terms of an anomalous diffusion of fluctuating Cooper pairs, that display 
a {\em quasi-confinement} (i.e., a slowing down of their diffusive dynamics) on length scales shorter than the 
inhomogeneity identified by tunnelling experiments. Here we assume this anomalous diffusive behavior of fluctuating 
Cooper pairs and calculate the effect of these fluctuations on the electron density of states above $T_c$. We find 
that the density of states is substantially suppressed up to temperatures well above $T_c$. This behavior, which 
is closely reminiscent of a pseudogap, only arises from the anomalous diffusion of fluctuating Cooper pairs in the 
absence of stable preformed pairs, setting the stage for an intermediate behavior between the two common paradigms 
in the superconducting-insulator transition, namely the localisation of Cooper pairs (the so-called bosonic scenario) 
and the breaking of Cooper pairs into unpaired  electrons due to strong disorder (the so-called fermionic scenario). 
\end{abstract}

\maketitle
\section{Introduction}
\begin{figure}[h!]
\centering
\includegraphics[width=.85\columnwidth]{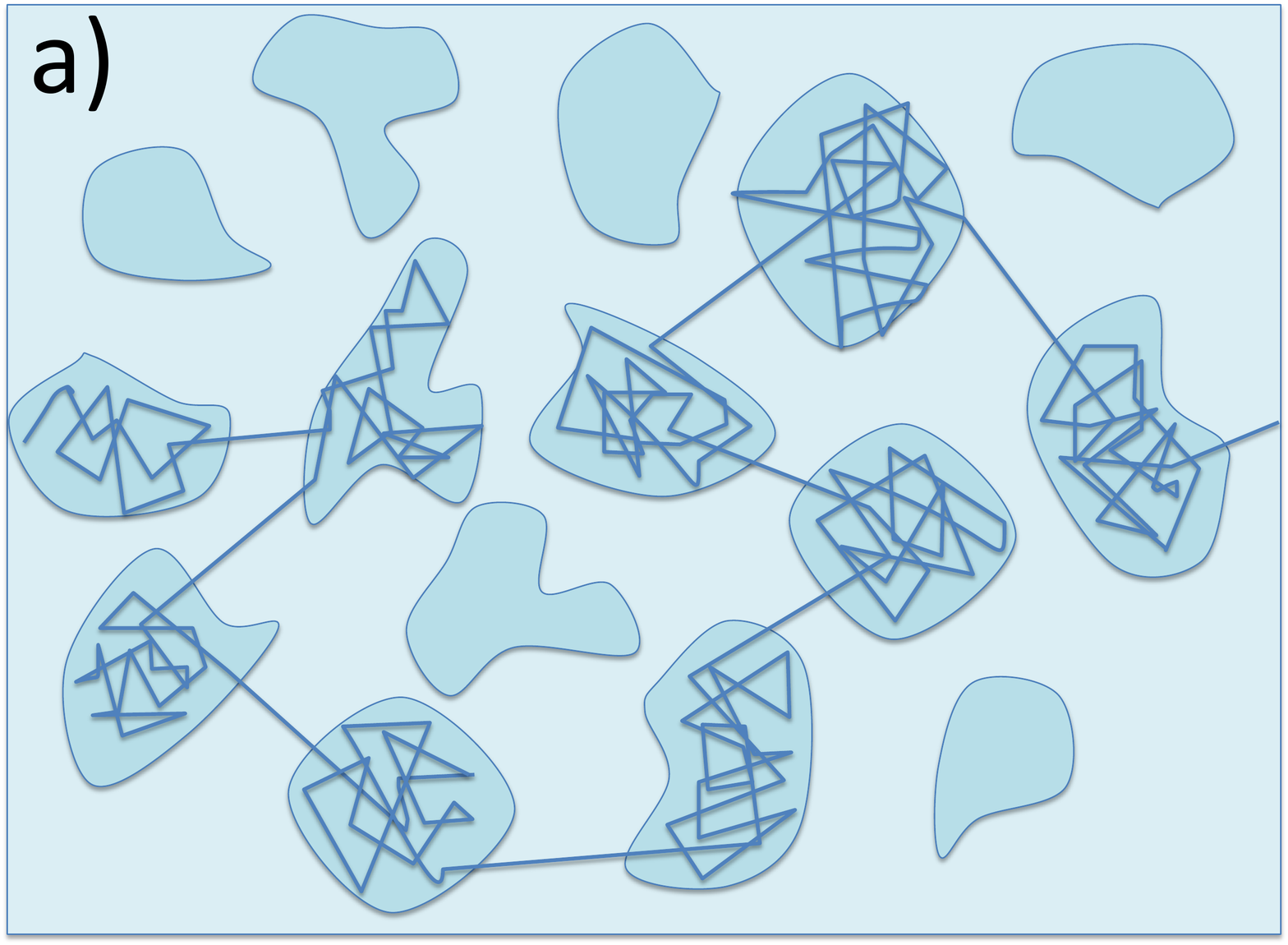}
\includegraphics[width=.85\columnwidth]{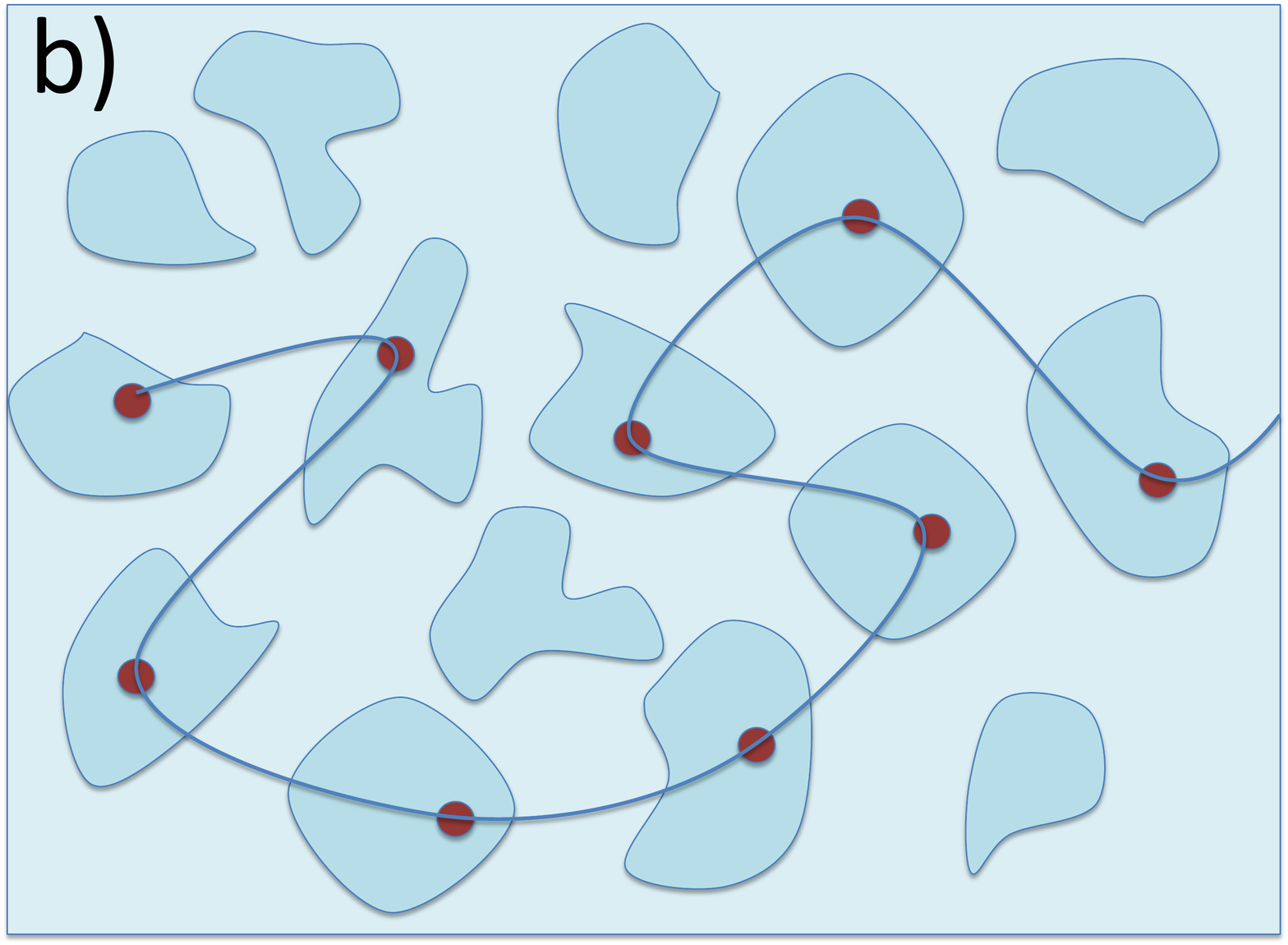}
\caption{Sketch of the anomalous diffusion of the fluctuating Cooper pairs induced by the emergent inhomogeneity in 
NbN thin films. (a) The fluctuating Cooper pairs above $T_c$ diffuse in the system, but are slowed inside the 
emergent inhomogeneities (represented by the darker regions). The multiple scattering inside these regions 
pictorially represents the diffusion slowdown leading to a {\em quasi-confinement} of the fluctuating Cooper pairs. 
This mechanism is expected to give rise to a pseudogap and to the zero-dimensional character of the 
paraconductive fluctuations. (b) Over longer distance and time scales a ``coarse-grained'' standard diffusive 
behaviour of the fluctuating Cooper pairs is recovered, leading to the usual two-dimensional AL behaviour of 
the paraconductivity.}
\label{fig:0D-2D}
\end{figure}

The physics of the Superconductor-Insulator Transition (SIT) in 
disordered superconducting thin films \cite{Goldman:1998jy} is attracting an ever increasing interest 
both for applicative purposes \cite{Goltsman:2001,Hofherr:2010i} and for fundamental reasons 
\cite{Dubi,Bard,Sacepe:2008jx}. While in structurally granular thin films the intrinsic granularity plays an evident 
role, the situation is more involved in nominally homogeneous (i.e., non-granular) disordered thin films. 
On the one hand, what is often referred to as the ``fermionic'' scenario proposes that the SIT is driven
by the reduced screening of the Coulomb repulsion with increasing disorder, that leads to a weakening of
pairing and to a reduction of the critical temperature $T_c$ \cite{FinkelShtein:1987}. In this case, the 
insulating state hosts localized fermions and standard paraconductive fluctuations are expected above $T_c$, 
due to Gaussian-distributed short-lived Cooper pairs. On the other hand, tightly-bound Cooper pairs survive the 
SIT in a ``bosonic'' scenario, in which the gap persists above $T_c$ despite the loss of phase coherence. In 
this framework, the bosonic pairs either localize because the disorder-enhanced Coulomb interaction destroys 
their phase-coherent motion at large scales \cite{Fisher:1990zza,Fisher:1990zz}, or disorder itself blurs the 
phase coherence without any relevant role of the Coulomb repulsion 
\cite{castellani:2012,Feigelman:2007bq,Feigelman:2010hp,Ioffe:2010dp}. 
In the latter case, it was also proposed that the superconducting state is characterized by an emergent 
disordered glassy phase \cite{Ioffe:2010dp}, with filamentary superconducting currents \cite{castellani:2012}.  
An anomalous distribution of the superconducting order parameter was proposed by theorists 
\cite{Ioffe:2010dp,Lemarie:2013bk}, and observed experimentally \cite{Sacepe:2008jx,Sacepe:2011jm}. 
A numerical approach to uniformly disordered superconductors \cite{Bouadim:2011hx} has also shown 
that there is a continuous evolution \cite{Trivedi:2012cj} from the weak-disorder limit, where the system has a rather 
homogeneous fermionic character, to the strong-disorder limit, where marked inhomogeneities appear in the 
superconducting order parameter, with an emergent bosonic nature characterized by a single-particle gap persisting 
on the insulating side of the SIT. A great deal of experimental activity has been devoted to this more disordered 
realization of the SIT \cite{Sacepe:2008jx,Kamlapure:2013kh}. The intermediate situation, where Cooper pairs 
begin to evolve into bosonic pairs, but keep their fermionic character, has been recently investigated by 
scanning tunnelling microscopy (STM) and transport measurements in NbN thin films. STM revealed the occurrence 
of an emergent inhomogeneous state and pseudogap effects over scales of a few tens of nanometers \cite{Carbillet}.
This intermediate-scale inhomogeneity affected the transport properties with the paraconductivity 
displaying a crossover from a seemingly zero-dimensional \cite{Varlamov-Larkin,varlamov-rmp} 
Aslasmazov-Larkin behaviour to the 
expected two-dimensional 
behaviour when $T_c$ was approached from above. This crossover was interpreted in terms of an anomalous slowing 
down of the dynamics of fluctuating Cooper pairs at length scales smaller than those set by the emergent 
inhomogeneity. Fig.\,\ref{fig:0D-2D} pictorially illustrates this effect.

In this paper we explore other consequences of this intriguing emergence of inhomogeneity. Specifically, we 
consider the effects of this anomalous dynamics of the fluctuating Cooper pairs on the electron density of states 
above $T_c$, showing that, in the presence of disorder, the slowing down of Cooper pair fluctuations
may give rise to pseudogap effects on the same length scale over which the Cooper pairs are {\em quasi-confined}. 
This will explain the occurrence of the pseudogap observed in STM experiments despite the persistence of purely 
fluctuating Cooper pairs (as opposed to the stable pairs of the bosonic scenario), which are 
needed to account for the standard Aslamazov-Larkin paraconductivity 

The structure of this paper is as follows. In the Sec.\,\ref{two} we present our phenomenological scheme 
involving the anomalous diffusion of fluctuating Cooper pairs. Sec.\,\ref{three} is devoted to the theoretical 
many-body calculation that determines the effects of the fluctuating Cooper pairs on the electron 
density of states (DOS). This section will also present the systematic analysis of these effects and a comparison 
with the experimental results of STM on NbN. Sec.\,\ref{four} contains our final remarks and conclusions.

\section{Theoretical background}
\label{two}

The core idea of the phenomenological theoretical framework used in Ref.\,\onlinecite{Carbillet} 
to fit the paraconductivity of NbN thin film is that there exist some regions, inside the superconducting 
film, where the lifetime of fluctuating Cooper pairs above the superconducting critical temperature $T_c$ is longer
than expected for a standard diffusion process, resulting in a \textit{quasi-confinement} of the fluctuating Cooper 
pairs within nanoscopic regions, dubbed supergrains, to emphasize the fact that they do not correspond to the 
structural grains. Indeed, these regions, of a linear dimension $L_i\approx 50$\,nm, are way larger than the 
typical disorder length scale of a few nanometers structurally present in nominally homogeneous NbN films. In this 
scenario, fluctuations with characteristic wavelength smaller than the typical inhomogeneity dimension $L_i$ are
more long-lived than they would be in a standard diffusion process, thus 
giving an explanation to the experimentally observed paraconductivity anomalies, resembling a zero-dimensional 
behaviour.

It is well established from the theory of fluctuations in superconductors, that the propagator 
of the fluctuating Cooper pairs in the weak-coupling (BCS) limit has the following form 
\cite{Varlamov-Larkin,varlamov-rmp}
\begin{equation}
\label{eq:Prop}
L(\vec{q},\Omega)\simeq -\frac{8T}{N_0}\frac{1}{\tau_{GL}^{-1}+\varepsilon(\qvec)-i\Omega},
\end{equation}
where $N_0$ is the electron DOS at the Fermi level, $\tau_{GL}^{-1}=\frac{8T_c}{\pi}\log\frac{T}{T_c}$ is the 
inverse Ginzburg-Landau lifetime, and 
$\varepsilon(\qvec)$ is the dispersion law, determining the increase of the inverse diffusion time with 
increasing magnitude of the wave vector $\qvec$ of the fluctuating mode. In this work we adopt units such that 
the Planck constant $\hbar$ and the Boltzmann constant $k_B$ are unity.

\begin{figure}[ht]
\centering
\includegraphics[width=.85\columnwidth]{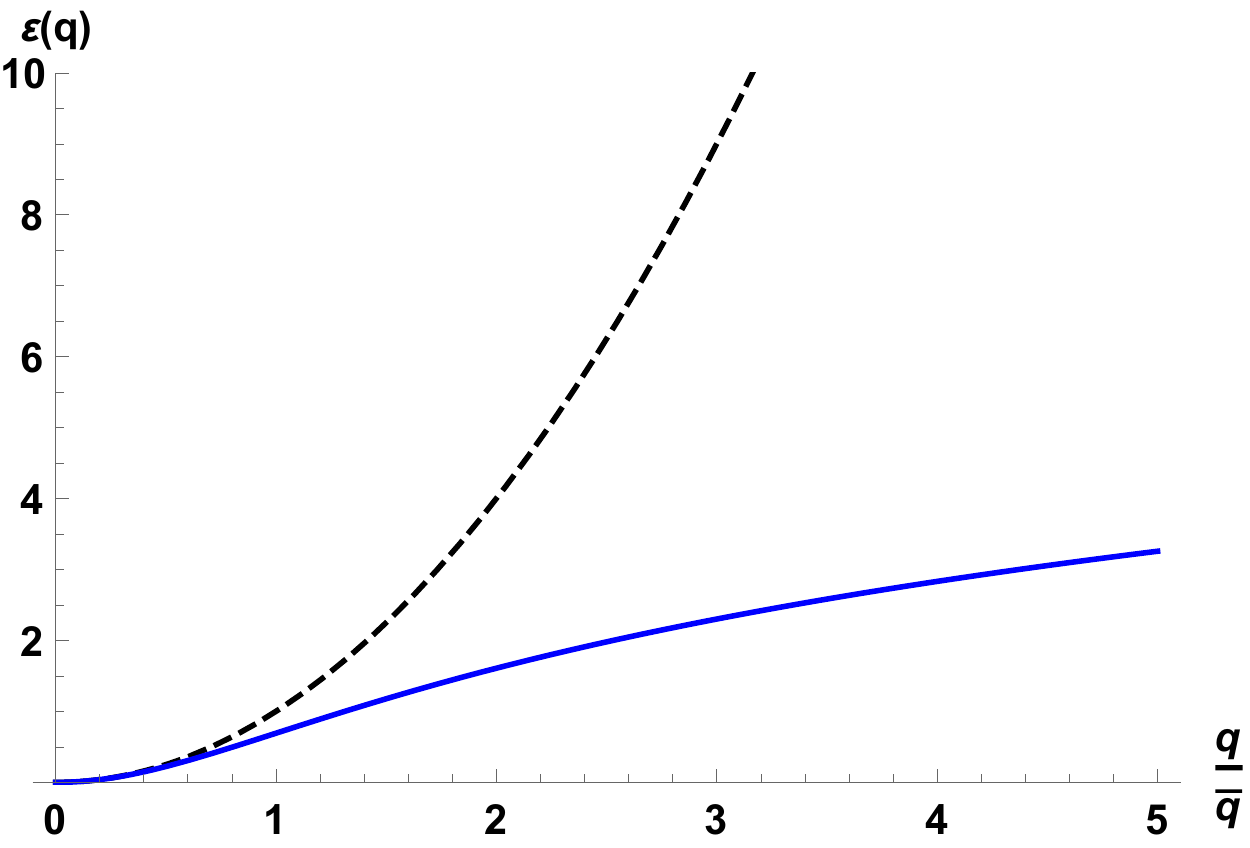}
\caption{The two different diffusion laws: the dashed line represents the standard quadratic law, while the 
blue curve is the anomalous one. It is possible to observe that on small momenta the two coincide asymptotically, 
while above $q/\bar{q}=1$ the anomalous diffusion becomes quite smaller than the standard one, representing the 
quasi-confinement of the fluctuating Cooper pairs inside the supergrains.}
\label{fig:Diff}
\end{figure}

The standard results from fluctuation 
theory are obtained with a quadratic diffusion law $\varepsilon(\qvec)=Dq^2$ (where $q\equiv |\qvec|$). 
On the other hand, the anomalous diffusion should give higher lifetimes for wavelengths 
smaller than $L_i$ and recover the standard behaviour at larger scales. It was found that a good description of 
experimental paraconductivity measurements on NbN thin films was achieved through the following anomalous 
diffusive expression
\be
\label{Eq:anomalous Diff}
\varepsilon(\vec{q})=D\bar{q}^2\log\left(1+\frac{q^2}{\bar{q}^2}\right),
\ee
where $D$ is the diffusion constant and $\bar{q}\approx L_i^{-1}$. Fig.\, \ref{fig:Diff} compares the 
standard quadratic diffusion law (black dashed line) with the anomalous diffusion of Eq.\,(\ref{Eq:anomalous Diff})  
(solid blue line). We point out that the value of the diffusion constant $D$ in the presence of disorder
may be severely suppressed with respect to the standard BCS value \cite{Varlamov-Larkin,varlamov-rmp}. 
As long as $D$ stays 
finite, however, its value does not appear in the AL paraconductivity in two dimensions, as the consequence of 
a cancellation enforced by gauge invariance \cite{caprara2009}. Disorder may also introduce corrections to the BCS value 
of $\tau_{GL}^{-1}$, as it seems indeed to be the case in the more strongly disordered NbN films \cite{Carbillet}.

At small momenta the two lines coincide since the system is normally diffusive over large length scales.
At $q\geq\bar{q}$ instead, the anomalous diffusion becomes much smaller than the quadratic one, indicating that 
for these modes the diffusion occurs with characteristic frequencies
$\Omega \sim \tau_{GL}^{-1}+\varepsilon(\qvec)$ much smaller (i.e., with much longer characteristic times) than 
in the standard case $\Omega \sim \tau_{GL}^{-1}+Dq^2$

The present work aims at understanding the effects of such an anomalous diffusion of fluctuating Cooper pairs 
on the electron density of states of a two-dimensional superconductor.

\section{Density of States from anomalous diffusion of fluctuating Cooper pairs}
\label{three}

Following a standard quantum many-body approach, the density of states of an electrons system is 
given by the imaginary part of its retarded Green function $\mathcal{G}^R(\vec{k},\omega)$:
\begin{equation}
\label{eq:N}
N(\omega)=-\frac{1}{2\pi}\int \frac{\mathrm d^2\kvec}{(2\pi)^2}\Im m\,[\mathcal{G}^R(\vec{k},\omega)].
\end{equation}

\begin{figure}[h!]
\includegraphics[width=\columnwidth]{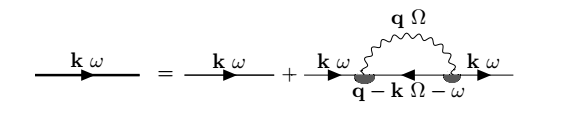}
\caption{Feynman diagram for the contribution of fluctuating Copper pairs to the 
electron Green function. The thick/thin solid line represents the dressed/bare electron Green function, the 
wavy line represent the propagator of fluctuating Cooper pair, and the shaded semicircles represent the impurity 
ladders that dress the electron-Cooper pair fluctuation vertex and embody the effect of microscopic disorder.}
\label{Fig:Feynman}
\end{figure}

The Feynman diagrams of figure \ref{Fig:Feynman} give the fluctuative contribution to the Green function in the 
dirty limit \cite{Abrahams,Pseudogap,DiCastro}. The wavy line corresponds to the propagator of fluctuating Cooper pairs 
$L(\vec{q},\Omega_n)$. While here  we adopt the finite-temperature formalism with Matsubara frequencies 
$\Omega_n$ \cite{Varlamov-Larkin,varlamov-rmp}, the form in Eq.\,(\ref{eq:Prop}) has been analytically continued to 
real frequency 
$\Omega$ in order to obtain a diffusive pole for the fluctuating Cooper pairs above $T_c$. The shaded semicircles are 
vertices coupling the Cooper pair fluctuations with the electron quasiparticles. In the presence of disorder due to 
quenched impurities and in the so-called ladder approximation \cite{Varlamov-Larkin,varlamov-rmp} they read
\begin{equation}
\label{Eq:Lambda}
\Lambda(\vec{q},\omega_m,\Omega_n)=\frac{1}{\tau}\frac{1}{\varepsilon(\vec{q})+
\left|2\omega_m+\frac{\mathrm{sign}\,(\omega_m)}{\tau}-\Omega_n\right|},
\end{equation}
where $\tau$ is the relaxation time of the electronic scattering on the impurities, while $\omega_m=(2m+1)\pi T$ 
and $\Omega_n=2n\pi T$ are the fermion and boson Matsubara frequencies, respectively.

From the second diagram in the r.h.s. of Fig.\,\ref{Fig:Feynman}, we obtain the corrections to the electron Green 
function due to fluctuating Cooper pairs. Upon integration over the electron momenta $\kvec$, we obtain
\begin{equation}
\label{Eq:dN}
\begin{split}
\delta G(\omega_m)=&-\frac{T}{\pi N_0}\Bigr[\sum_{\Omega_n}\int\frac{\mathrm d^2\qvec}{(2\pi)^2}L(\vec{q},\Omega_n)
\Lambda^2(\vec{q},\omega_m,\Omega_n)\\
&\int\frac{\mathrm d^2\kvec}{(2\pi)^2}G_0^2(\vec{k},\omega_m)G_0(\vec{q}-\vec{k},\Omega_n-\omega_m)\Bigl],\\
\end{split}
\end{equation}
where $G_0(\vec{k},\omega_m)=[i\omega_m-v_F(k-k_F)]^{-1}$ is the free-electron Green function in the Matsubara 
formalism, $v_F$ and $k_F$ being the Fermi velocity and Fermi wave vector, respectively. The sum over boson 
Matsubara frequencies and the integral over fermion momenta can be carried 
out analytically as it was done in Ref.\,\onlinecite{Pseudogap} for the case of standard diffusion of 
fluctuating Cooper pairs. Analytically continuing the fermion frequencies ($i \omega_m \to \omega-i0^+$) and taking 
the imaginary part in Eq.\,(\ref{Eq:dN}), one obtains the variation of the DOS 
\begin{equation}
\label{eq:result}
\begin{split}
\delta N(\omega)&=-\frac{16T^2}{\pi N_0\tau}\int\frac{q\,\mathrm d q}{\left[\tau_{GL}^{-1}+
\varepsilon(q)\right]\left[\tau_{GL}^{-1}+2\varepsilon(q)\right]^2}\\
&\times\frac{\left(4|\omega|+\frac{1}{2\tau}\right)^2-
\left[\tau_{GL}^{-1}+\varepsilon(q)\right]^2}{\left\{\left(4|\omega|+\frac{1}{2\tau}\right)^2
+\left[\tau_{GL}^{-1}+\varepsilon(q)\right]^2\right\}^2},
\end{split}
\end{equation}
where the last integration over $q=|\qvec|$ can be carried out numerically, and we made explicit that
$\varepsilon(\qvec)$ only depends on $q$.

Starting from the numerical evaluation of Eq.\,(\ref{eq:result}), we systematically investigate the role of 
disorder (namely of the scattering time $\tau$) and of the inhomogeneity length scale ($L_i$). 

One first generic finding is that the anomalous diffusion of fluctuating Cooper pairs substantially increases the 
size and the extension in temperature above $T_c$ of the pseudogap effects (i.e., of the partial DOS suppression 
that occurs above $T_c$ because of superconducting fluctuations) in comparison with the standard diffusion law 
$\varepsilon(q)=Dq^2$. These effects are markedly larger in the disordered case than in the clean case 
(i.e., in the $\tau\to \infty$ limit): Only in the disordered case the DOS suppression can
become quantitatively comparable to the suppression observed by STM experiments in NbN. This is why in the 
following we will focus on the disordered case only.

\begin{figure}[h!]
\centering
\includegraphics[width=.5\textwidth]{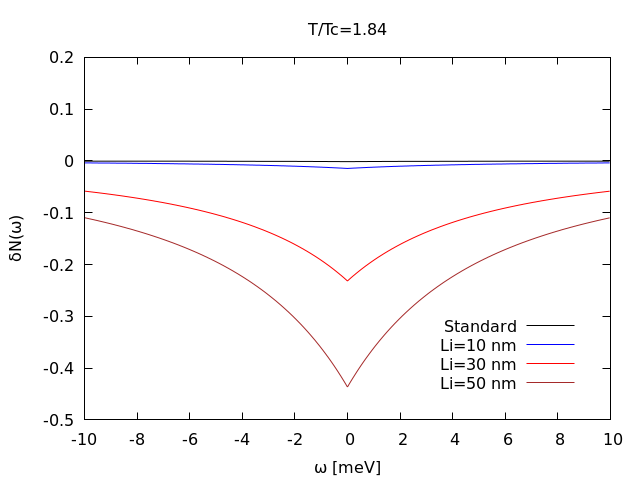}
\caption{Effects of (anomalously diffusing) Cooper pair fluctuations on the quasiparticle DOS at $T/T_c=1.84$,
at fixed level of disorder.  
The elastic scattering rate is $\frac{1}{\tau}=20$\,meV\,$\approx 0.067\,E_F$, where
$E_F=v_F k_F$ is the Fermi energy. The black line shows the result for standard diffusion, with barely any visible 
effect as compared to the much stronger suppression observed in the case of anomalous diffusion.}
\label{fig:comparison}
\end{figure}

In Fig.\,\ref{fig:comparison}, different $\delta N(\omega)$ curves are compared, at $T/T_c=1.84$, rather far above 
$T_c$. The black line represents the curve obtained with the standard diffusion law, while the coloured ones have 
been obtained with different  values of $L_i=20,30,50$\,nm, [i.e. different $\overline{q}$ in 
Eq.\,(\ref{Eq:anomalous Diff})], all in the range of the inhomogeneity sizes observed in 
Ref.\,\onlinecite{Carbillet}.

It is evident that anomalous diffusion greatly enhances the pseudogap effects. Quite far above $T_c$, 
at $T/T_c=1.84$, the DOS suppression in the standard diffusion case (black solid line) is barely appreciable 
adopting a common scale, being much smaller than the suppression observed in the case of anomalous diffusion. 
This demonstrates that these pseudogap 
effects are quite robust in temperature. We also notice that increasing the parameter $L_i$ we obtain a 
more pronounced DOS suppression, as this reduces $\bar q$, thus extending the region in $q$ space where 
the anomalous diffusion of fluctuating Cooper pairs occurs.

\begin{figure}[h!]
\centering
\includegraphics[width=.5\textwidth]{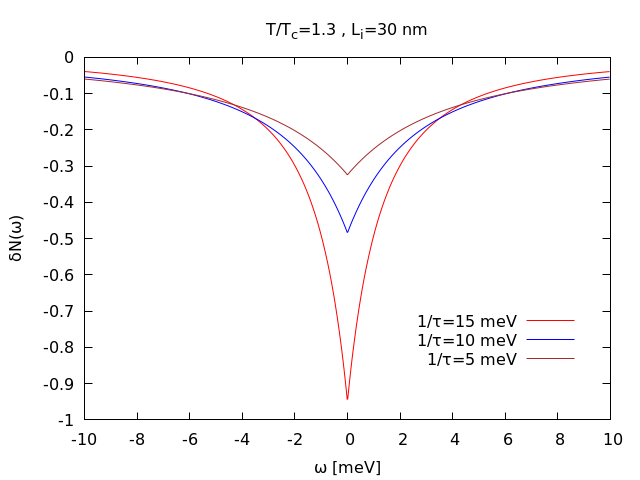}
\caption{Effect of the disorder strength on the Cooper fluctuation contribution to the DOS at $T/T_c=1.3$ and 
$L_i=30$\,nm.}
\label{Fig:tau}
\end{figure}

In Fig.\,\ref{Fig:tau}, we report the results of our analysis on the effects of disorder. It is evident that, 
for a given fixed value of $L_i=30$\,nm,  at $T/T_c=1.3$, the DOS suppression increases by increasing the 
elastic scattering rate $\frac{1}{\tau}$. This indicates that the homogeneously distributed microscopic disorder 
represented by the impurities is cooperative with the large-scale inhomogeneity producing a stronger effect of the 
anomalous diffusion of Cooper pair fluctuations.

The results reported above show that the pseudogap effects induced by the anomalous diffusion of fluctuating 
Cooper pairs are large enough to account for the sizable pseudogap effects observed by STM in NbN. On the other hand,  
the DOS corrections shown in the Figs.\,\ref{fig:comparison}  and \ref{Fig:tau} systematically display a 
suppression (i.e., a negative correction) with respect to the reference DOS of the normal metallic state at 
high temperature. Therefore, result cannot account for the experimental observation of coherence effects that 
rather symmetrically produce an enhancement of DOS at finite energy (finite bias in tunnel experiments) above 
or below the Fermi level ($\omega=0$). Of course the spectral weight is not lost in our calculations, but is 
simply distributed over a very broad range, larger than the range of the figures. To obtain more realistic tunnel 
spectra, it is crucial to recall that in the transport experiments of Ref.\,\onlinecite{Carbillet}, 
the transition temperature in the two-dimensional regime that sets in when the true $T_c$ is approached is larger 
than the critical temperature of the anomalously diffusing fluctuating Cooper pairs that shape the 
paraconductivity data at higher temperatures further away from $T_c$, in the regime where fluctuations have the
seemingly zero-dimensional character. This observation, substantiated in Fig. 4b of Ref.\,\onlinecite{Carbillet}, is 
not surprising, because the 0D regime occurs at higher temperatures and is dominated by the anomalous diffusion 
of fluctuating Cooper pairs over length scales shorter that $L_i$ and at these scales the systems is seemingly
unaware of the actual critical temperature $T_c$ at which the true two-dimensional global superconducting state 
is established.
 
\begin{figure}[h!]
\centering
\includegraphics[width=.5\textwidth]{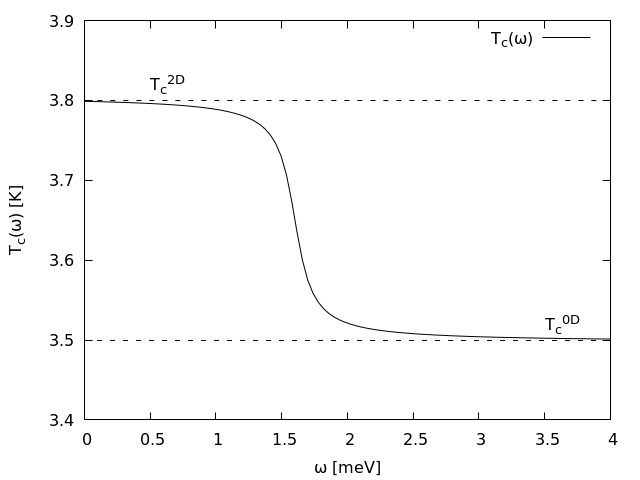}
\caption{The curve representing the proposed $T_c(\omega)$. A smooth change from the two-dimensional (2D) 
regime to the zero-dimensional (0D) one can be tuned using an arctangent interpolating function. The values
of the asymptotic critical temperatures are those obtained in Ref.\,\onlinecite{Carbillet}}
\label{fig:Tc}
\end{figure}

Therefore, in order to mimic  this `flowing' of the critical temperature when shorter and shorter length scales 
(i.e., larger and larger inverse time scales) dominate the Cooper fluctuations, we assumed a smooth
step-like energy dependent $T_c(\omega)$, as shown in Fig.\,\ref{fig:Tc}.

\begin{figure}[h!]
\centering
\includegraphics[width=.48\textwidth]{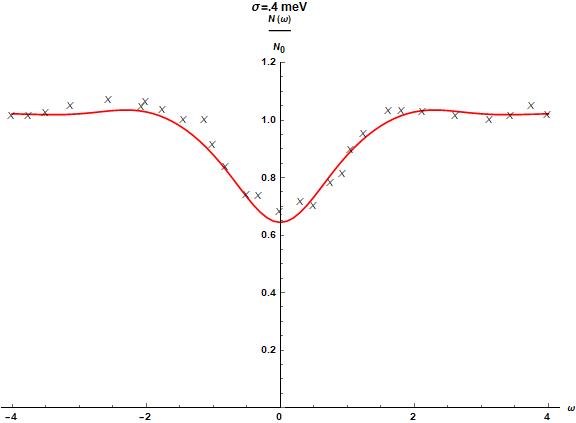}
\caption{Comparison of the theoretical curve (red line) with experimental data ($\times$) 
at a temperature $T=1.1\,T_c$ and with $L_i=16.5$\,nm as typical size of the inhomogeneity. 
Here $1/\tau=5$\,meV. The theoretical curve has been convoluted with a gaussian with variance 
$\sigma=0.4$\, meV to account for the experimental resolution}
\label{fig:res}
\end{figure}

With this phenomenological assumption, we then obtain the fitting of tunnel spectrum reported in 
Fig.\,\ref{fig:res} for $T=1.1\,T_c$. The data were obtained taking the experimental spectra 
of the large pseudogap regions (the supergrains) and dividing them by the corresponding measurements in the
small pseudogap regions, to extract the contribution of anomalously diffusing Cooper pair fluctuations
and get rid of any background contribution. 
We used $L_i$ and $\tau$ as adjustable parameters. The fitting DOS suppression (normalised to the DOS in the 
metallic state) was obtained for $L_i=16.5$\,nm, quite comparable with the typical size of supergrains observed in 
NbN films, and for $1/\tau=5$\,meV, for which we have no independent determination. Fig. \ref{fig:res} reports the
comparison between the theoretical calculations and the experimental data. The theoretical curve has  also been convoluted with a gaussian with variance 
$\sigma=0.4$\, meV to account for the experimental resolution. 

The agreement between the theoretical curve and experimental data is quite satisfactory at every $\omega$. 
The convolution with the finite-resolution gaussian plays a role ony near $\omega=0$, where it transforms the cusp-like
theoretical curves (see, e.g., Fig. \ref{Fig:tau}) in the more rounded red line of Fig. \ref{fig:res}.

\section{Conclusions and outlooks}
\label{four}

In this paper, we presented a theory for the DOS suppression due to anomalously diffusing fluctuating Cooper pairs
above the critical temperature $T_c$ of a two-dimensional superconductor. Our theoretical results highlight 
the effectiveness of the anomalous diffusion of fluctuating Cooper pairs in enhancing the pseudogap at temperatures 
well above $T_c$. Since a similar effect was observed experimentally in some ultrathin NbN films, we compared our 
curves with STM experiments, in order to get an estimate of the parameters of the theory.
Good agreement with experimental data is achieved, especially after introducing the natural idea that Cooper fluctuations
at different energies may `perceive' different critical temperatures. This led us to the phenomenological assumption that
the critical temperature should, in our theory, be frequency dependent, as a consequence of the 
0D-2D crossover, with the asymptotic high-frequency (0D) and 
low-frequency (2D) values fitted to match the experimental paraconductivity \cite{Carbillet}, and a smooth 
interpolating behaviour at intermediate-frequency.
This work has been carried out on a phenomenological basis in the form of the anomalous diffusive law of Cooper pairs and of $T_c(\omega)$ 
and the understanding of the actual microscopical reasons giving rise to the inhomogeneities, their length scales and their effects on the
above laws is lacking.
Nevertheless, we point out that, our work opens the way to a new microscopic understanding of
the SIT, in which Cooper pairs are not stable bosonic entities, but keep a fluctuating character. It is their slowing down that produces 
marked pseudogap effects mimicking that of preformed pairs and strengthening the idea of a gradual and continuous evolution
between the bosonic and fermionic scenarios.
\\
~
\\
~
\par\noindent
{\bf Acknowledgments.} S.C. thanks all the colleagues of the ESPCI in Paris for their warm hospitality and for many 
useful discussion while part of this work was done. S.C. and M.G. acknowledge financial support of the University of Rome 
Sapienza, under the Ateneo 2017 (prot. RM11715C642E8370) and Ateneo 2018 (prot. RM11816431DBA5AF projects. Part of 
the work was supported through the Chaire Joliot at ESPCI Paris. This work was also supported by EU through the COST 
action CA16218 NanocoHybri.

\end{document}